\documentclass[aps,prl,twocolumn,superscriptaddress,nofootinbib,floatfix,longbibliography]{revtex4-2}

\usepackage{graphicx}
\usepackage{amsmath}
\usepackage{amssymb}
\usepackage{bm}
\usepackage[colorlinks=true,citecolor=blue,linkcolor=blue,urlcolor=magenta]{hyperref}

\newcommand{\keV}{\ensuremath{\,\mathrm{keV}}}
\newcommand{\MeV}{\ensuremath{\,\mathrm{MeV}}}
\newcommand{\GeV}{\ensuremath{\,\mathrm{GeV}}}
\newcommand{\cms}{\ensuremath{\,\mathrm{cm^2}}}
\newcommand{\ER}{\ensuremath{E_R}}
\newcommand{\mchi}{\ensuremath{m_\chi}}
\newcommand{\vmin}{\ensuremath{v_{\rm min}}}

\begin{document}

\title{Exothermic and Endothermic  Inelastic Dark Matter Interpretations at LZ: \\
Sideband Constraints and Future Prospects}

\author{James B.~Dent} 
\email{jbdent@shsu.edu}
\affiliation{Department of Physics and Astronomy$,$~ Sam ~Houston~ State~ University$,$~ Huntsville$,$~ TX~ 77341$,$~ USA}

\author{Jayden L.~Newstead}
\email{jnewstead@unimelb.edu.au}
\affiliation{ARC Centre of Excellence for Dark Matter Particle Physics$,$ \\~School of Physics$,$~ The~ University~ of~ Melbourne$,$~ Victoria~ 3010$,$~ Australia}

\begin{abstract}
The LZ experiment has extended its nuclear-recoil search to $270\keV$ and reports a single nuclear-recoil-like event, LZ230616, at $\sim250$~keV, in a region with negligible expected background. We show that exothermic inelastic dark matter, in which the ambient dark state down-scatters and releases its mass splitting $|\delta|$ as recoil energy, can naturally produce a recoil peak at this energy while remaining consistent with the null result at lower energies. Our simple fit leaves the mass unconstrained, because a heavy candidate can hide its peak above the LZ region-of-interest. However, we show that the empty high-energy sideband, which LZ uses for background validation, disfavors broadly peaked spectra. For an exothermic explanation, the sharper peaks produced by lighter dark matter with larger mass splittings are therefore preferred. Realizing this scenario requires cosmologically stable states and a leptophobic mediator. Normalizing to the one observed event, we predict $\simeq4.5$ signal events in LZ's projected 1000 live-day exposure, and show that existing XENONnT and PandaX-4T data could already test the interpretation in an extended analysis window.
\end{abstract}

\maketitle

Dual-phase xenon time projection chambers set the leading constraints on the spin-independent scattering of weak-scale dark matter (DM) off nuclei~\cite{LZ:2024zvo,XENON:2025vwd,PandaX:2024qfu}. These searches are optimized for recoil energies $\ER \lesssim 55\keV$, where one expects the nuclear scattering signal of a standard weakly interacting massive particle (WIMP) candidate. This is predicated on the assumption that the scattering is both elastic and momentum independent. Previous work highlighted the importance of extending this energy window to ensure that more exotic WIMP candidates would not be missed~\cite{Bramante:2016rdh,Barello:2014uda}, motivating analyses over a wider range of recoil energies~\cite{LZ:2023lvz,PhysRevD.109.112017}. Recently, LZ reported such a search, extending to $270\keV$, using $2.84$ tonne-years of exposure~\cite{LZ:2026axp}. This analysis detected a single signal-like event, LZ230616, with reconstructed energy $248\pm23_{\rm stat}\pm23_{\rm sys}\keV$ and an integrated background expectation of $0.0106\pm0.0008$ events. The collaboration quotes a local significance of $3.4\sigma$, and a global significance of $2.6\sigma$ depending on the fitted model.

Elastic scattering of halo dark matter is a poor explanation of an event at this energy. It is not excluded kinematically since the maximum recoil energy, $\ER^{\rm max} = 2\mu_{\chi N}^2 v^2_{\rm max}/m_N$, can exceed $248\keV$ for $\mchi \gtrsim 75\GeV$. However, standard WIMP elastic scattering spectra typically fall by four to six orders of magnitude between $\mathcal{O}(10\keV)$ and $\mathcal{O}(100\keV)$. This is driven by both the velocity distribution and the nuclear form factor. Thus, a standard WIMP producing a single event at $\sim250\keV$ is extremely improbable. To explain the event as a DM signal, the assumption of elastic or momentum independent scattering must be dropped. This was demonstrated in the LZ analysis itself, which finds $0.0\sigma$ for the elastic, momentum-independent limit of the isoscalar $\mathcal{O}_1$ operator, but up to $3.4\sigma$ for momentum-dependent operators~\cite{LZ:2026axp}.

Inelastic dark matter, in which the ambient dark state belongs to a nearly degenerate pair with mass splitting $\delta$, is a long standing extension to standard direct detection phenomenology~\cite{Tucker-Smith:2001myb,Tucker-Smith:2004mxa,Graham:2010ca}. The endothermic case, where $\chi$ up-scatters to the heavier $\chi^*$ (consuming kinetic energy), was considered by LZ over $0 < \delta < 350\keV$~\cite{LZ:2026axp}. Early interpretations of LZ230616 have followed this endothermic route, favoring TeV-scale candidates, notably a $\sim1\,$TeV Higgsino with $\delta \simeq 350\keV$~\cite{Su:2026rwz,Freese:2026sga,Wu:2026nhi,DiMauro:2026ldr} (though, see~\cite{Pospelov:2026ewn} for difficulties with this interpretation).\footnote{Other possibilities include absorption of a $\sim250\MeV$ fermionic state~\cite{Lou:2026idn}, a $Z_2$ Higgs partner~\cite{Nomura:2026qyq}, Peccei-Quinn origination~\cite{Visinelli:2026kgt,Yin:2026jnn}, and dark photon DM~\cite{Yamashita:2026ump}.}  

The opposite, exothermic case, with $\delta<0$, has yet to be considered as a possible explanation of the LZ event. In this scenario, the cosmologically abundant state is the heavier one, and it can undergo the down-scattering process $\chi^{*}N \to \chi N$, thereby releasing the mass splitting energy into the recoil products~\cite{Graham:2010ca,Batell:2009vb,Essig:2009nc}. When the mass difference is large compared to the incoming kinetic energy, the recoil spectrum can be quasi-monoenergetic, and its position is above the elastic endpoint. In this Letter we ask what masses and splittings could explain LZ230616, and show that the data, in particular the absence of events in LZ's high-energy sideband, favor a sharp peak. This is in contrast with existing endothermic inelastic explanations, which have broader peaks. Future analyses with a modest increase in data will be able to distinguish these two competing explanations.

\emph{Exothermic kinematics ---}
We consider two states $\chi$ and $\chi^{*}$ with $\delta \equiv m_\chi - m_{\chi^{*}} < 0$, coupled to the Standard Model through an off-diagonal current so that elastic scattering is absent at tree level. A fraction $f_* \equiv \rho_{\chi^*}/\rho_{\rm DM}$ of the halo must reside in $\chi^*$; we return to this below. Energy and momentum conservation give the minimum speed to deposit $\ER$ on a nucleus of mass $m_N$,
\begin{equation}
  \vmin(\ER) \;=\;
  \frac{1}{\sqrt{2 m_N \ER}}\,
  \left| \frac{m_N \ER}{\mu_{\chi N}} + \delta \right| \, .
  \label{eq:vmin}
\end{equation}
For $\delta<0$, this produces a peak at the energy
\begin{equation}
  E_0 \;=\; \frac{\mu_{\chi N}\,|\delta|}{m_N} \, ,
  \label{eq:E0}
\end{equation}
with half-width
\begin{equation}
  \Delta E \simeq
  \frac{\mu_{\chi N}\, v_{\mathrm{max}}}{m_N}\sqrt{2 m_N E_0},
  \label{eq:width}
\end{equation}
where $v_{\rm max} = v_{\rm esc} + v_{\rm lab}$ is the maximum incoming DM velocity in terms of the local galactic escape velocity and lab velocity. Given the observation of a single event, its energy fixes the best-fit peak position and thus the product $\mu_{\chi N}|\delta|$, while the width depends on $\mu_{\chi N}$ alone, which is only mildly constrained in this context.

However, since the event may sit to either side of the peak, requiring a recoil $\ER$ to be accessible at some speed $v \le v_{\rm max}$ gives
\begin{equation}
  \left|\, |\delta| - \frac{m_N \ER}{\mu_{\chi N}} \,\right|
  \;\le\; v_{\rm max}\sqrt{2 m_N \ER} \;\simeq\; 0.65\MeV ,
  \label{eq:window}
\end{equation}

where the last equality takes $\ER = 248\keV$, and the target xenon nuclear mass for $m_N$. For the exothermic case, the splitting must be in the range $1.6$--$2.9\MeV$ at $\mchi = 15\GeV$ and $0.3$--$1.6\MeV$ at $40\GeV$. Above $\mchi \simeq 75\GeV$ the elastic endpoint reaches $\sim248\keV$ and endothermic mass splittings start to become accessible. We note that lighter DM with larger mass splittings produce a narrow peak, while heavier DM and smaller mass splittings broaden the peak. Thus, as more data is collected, the emergence of a peaked/broad signal will discriminate between these two cases.

\begin{figure}[t]
  \includegraphics[width=\columnwidth]{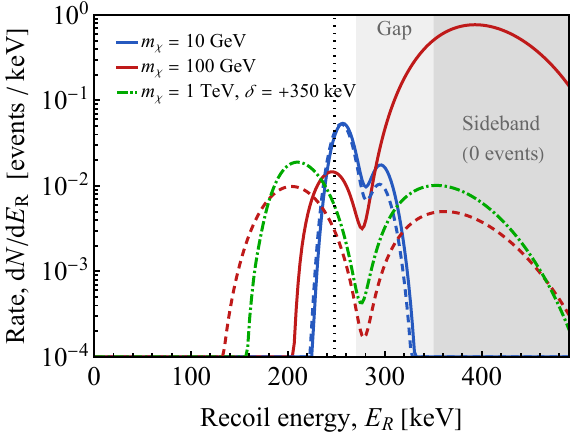}
  \caption{Recoil spectra for inelastic dark matter, each normalized to one expected event. Solid: we show the best-fit splitting for the given mass, using the search window alone. Dashed: we show the best-fit once the empty high-energy sideband (darker shading) is included in the likelihood. The heavy-mass spectrum shifts when including the sideband, while the light-mass spectrum is mostly unaffected. Green: the endothermic TeV benchmark of Ref.~\cite{Freese:2026sga} ($1\,$TeV, $\delta = +350\keV$), shown for comparison and likewise normalized to one event. The vertical dotted line denotes LZ230616.}
  \label{fig:spectrum}
\end{figure}

\begin{figure}[t]
  \includegraphics[width=\columnwidth]{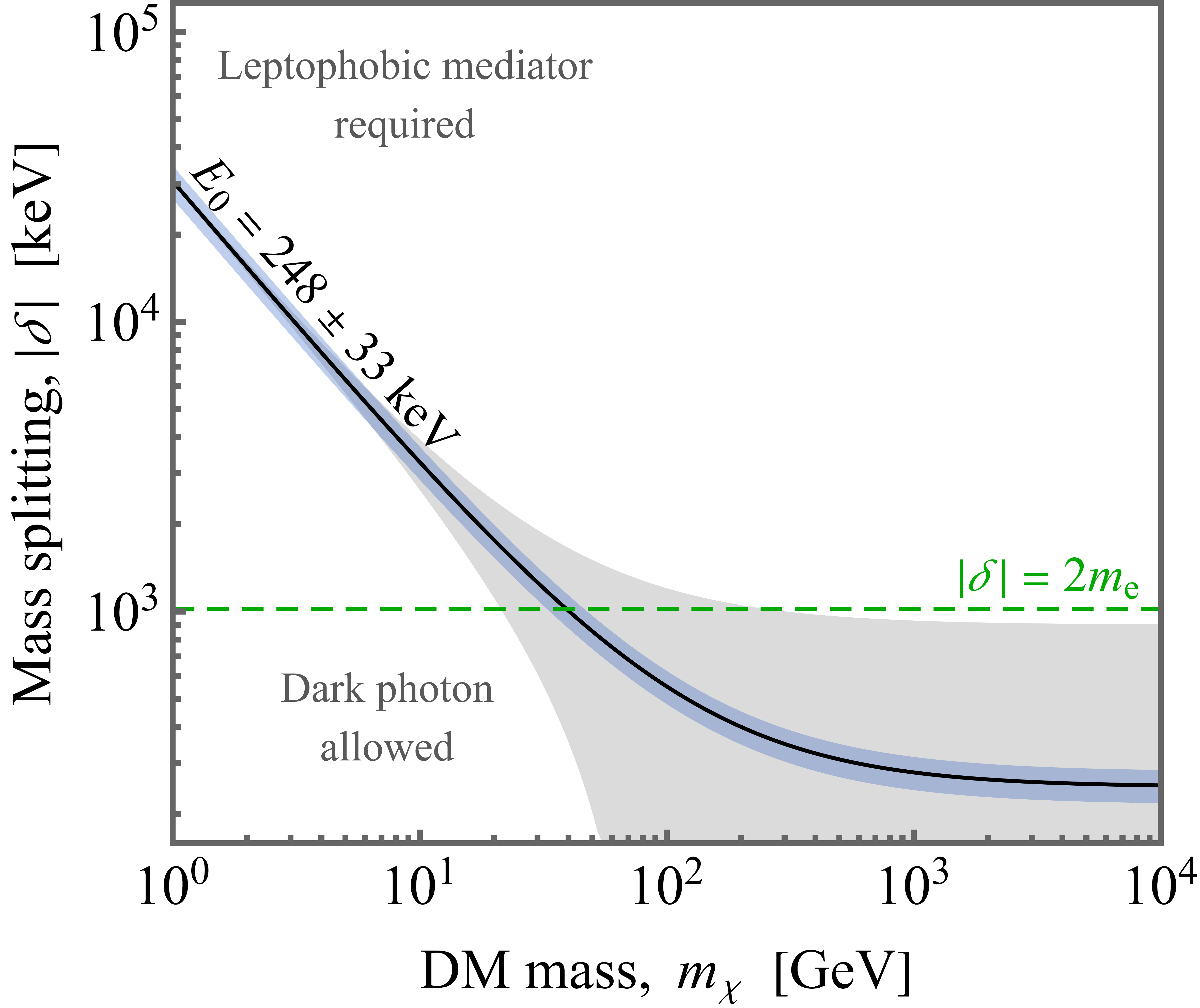}
  \caption{Favored exothermic parameter space for LZ230616. The black curve traces $|\delta| = E_0 m_N/\mu_{\chi N}$ with $E_0 = 248$ keV, while the blue region shows the $1\sigma$ error of the measured energy. The gray region is the full allowed kinematic window of Eq.~(\ref{eq:window}). Above the $|\delta| = 2m_e$ line the excited state decays to $e^+e^-$ for any electron-coupled mediator.}
  \label{fig:region}
\end{figure}

\emph{Fit to LZ230616 ---}
We fit the single observed event assuming spin-independent exothermic scattering, summed over xenon isotopes and using standard assumptions: a Helm form factor and the Standard Halo Model with conventions from Ref.~\cite{Baxter:2021pqo} (further details in the Supplemental Material). The fit is computed using an extended unbinned likelihood with LZ's published nuclear-recoil (NR) efficiency and assuming zero background. We profile over the cross section analytically, leaving a pure shape-based likelihood in $(\mchi,\delta)$. With a single event, the fit places the peak at the event energy, $E_0 \simeq 250$--$260\keV$, with $\sigma_p f_* \simeq (0.5$--$3)\times10^{-45}\cms$ across the allowed mass range, where $\sigma_p$ is the DM-proton cross section (and we assume no isospin violation). Example spectra are shown in Fig.~\ref{fig:spectrum}, while the allowed region in the $m_\chi$ vs. $\delta$ parameter space is given in Fig.~\ref{fig:region}. In Fig.~\ref{fig:region} we see that the allowed splitting falls roughly as $m_{\chi}^{-1}$  before crossing $|\delta| = 2m_e$ at $\simeq 40~{\rm GeV}$, a crossover point which determines the possible final states, thereby impacting DM model considerations as we will discuss below. For the region $m_\chi \gg m_N$, the splitting asymptotes to $E_0$.

Given that the single event occurs toward the edge of LZ's WIMP search region, the DM mass is poorly constrained. This is because a heavy candidate can evade the fit by placing $E_0$ above the $270\keV$ acceptance edge, leaving only the rising edge of the peak at the event energy (Fig.~\ref{fig:spectrum}). Thus, a fit using the LZ WIMP search region alone allows any mass between $1\GeV$ and $10\,$TeV. This mass region combined with the acceptance region is testable with more data. That said, to validate their background model, LZ defines a high-energy sideband at $800 < \mathrm{S1c} < 1700$~phd (roughly $350$--$680\keV$) and observes zero events there in the science sample~\cite{LZ:2026axp}. It follows that models that predict peaks above the LZ WIMP search region are actually strongly disfavored. For example, the expected sideband yield per observed event is $0.02$ at $\mchi = 10\GeV$, $19$ at $42\GeV$ and $73$ at $100\GeV$. Treating the sideband as a second, background-free signal region and assuming the published NR efficiency plateau extends into this region (see Supplemental material for more details), the fit then prefers $\mchi \lesssim 90\GeV$ at $95\%~{\rm CL}$.

Therefore the empty sideband prefers a light-mass peak over broad heavy spectra. Additionally, applying this to the endothermic interpretation: for the $1\,$TeV, $\delta = +350\keV$ benchmark of Ref.~\cite{Freese:2026sga}, which peaks at $\simeq210\keV$, one expects $0.6$ events in the sideband per search-window event (see Fig.~\ref{fig:spectrum}). This is consistent with LZ's zero observed events. The sideband therefore discriminates among exothermic candidates, but at the current exposures, it does not disfavor the endothermic solution. The interval $600$--$800$~phd ($\simeq270$--$350\keV$), between the two published regions, is not included in our fit and thus still allows heavier WIMPs with non-zero rate in this region. Future analyses with a wider NR range would allow for an improved analysis with more discriminating power.

\emph{Constraints on the excited state ---} The preceding discussion and spectral fits are independent of the underlying inelastic model. To ensure the exothermic scenario is viable, we must consider the lifetime of the excited state. Exothermic scattering can occur when some fraction, $f_*$, of the ambient DM is in the higher-mass state. However, the same coupling that produces a scattering signal makes $\chi^*$ unstable. Therefore, the $\chi^*$  state must either be stable on cosmological timescales or be constantly repopulated through astrophysical processes.  Given the large mass splittings required to fit the LZ signal, the $\chi^*$ cannot be easily populated by astrophysical processes and thus must be stable on cosmological time scales. Additionally, depending on the decay process, indirect detection can constrain the lifetime to be 6--10 orders of magnitude longer than the age of the universe~\cite{Essig:2013goa}.

A concrete realization of exothermic scattering involves a pseudo-Dirac pair and a heavy vector, kinetically mixed with the photon~\cite{Batell:2009vb,Finkbeiner:2009mi}. In our scenario, DM lighter than $\sim40$ GeV requires $|\delta| > 2m_e$. In this region, the decay is dominated by the process $\chi^{*}\to\chi e^+e^-$, with lifetime
\begin{equation}
  \tau_{\chi^{*}} \;\approx\; 4\times10^{9}\,\mathrm{s}\;
  \frac{1}{f_{\rm PS}}
  \left(\frac{\sigma_p}{10^{-45}\cms}\right)^{-1}
  \left(\frac{|\delta|}{2\MeV}\right)^{-5} ,
  \label{eq:lifetime}
\end{equation}
where $f_{\rm PS} \le 1$ is the $e^+e^-$ phase-space factor near threshold~\cite{Fitzpatrick:2021cij,CarrilloGonzalez:2021lxm}.  Taking $f_* = 1$ (to minimize the $\sigma_p$ allowed by the single LZ event, and thus maximise the lifetime), we have $\tau_{\chi^*} \sim 10^{9}\,$s for $\mchi = 15\GeV$, some eight orders of magnitude short of the age of the Universe. Therefore, the electron decay channel must be closed by considering $|\delta| < 2m_e$ (requiring $\mchi \gtrsim 40\GeV$), or a leptophobic mediator. Fig.~\ref{fig:region} displays the allowed kinematic window on both sides of this divide using Eq.~(\ref{eq:window}), with a much narrower band for the $|\delta| > 2m_e$ case when fit to the LZ event.

For the leptophobic mediator case, the leading hadronic decay channel would require $|\delta| > 2m_\pi$, which is far above the $|\delta|$ we require to fit the LZ signal. Below this threshold the lifetime is determined by the loop-induced process $\chi^{*}\to\chi\,3\gamma$, which is cosmologically slow for weak kinetic mixing~\cite{Finkbeiner:2009mi}.

The interpretation of the LZ signal in terms of exothermic scattering therefore presents two distinct possibilities: a $\sim40$--$90\GeV$ state with a sub-MeV splitting and a dark photon mediator, or a sub 40 GeV state with a larger $\sim1$--10 MeV splitting whose mediator must not couple to leptons (the latter case is mildly preferred by the empty sideband discussed previously).

\begin{figure}[t]
  \includegraphics[width=\columnwidth]{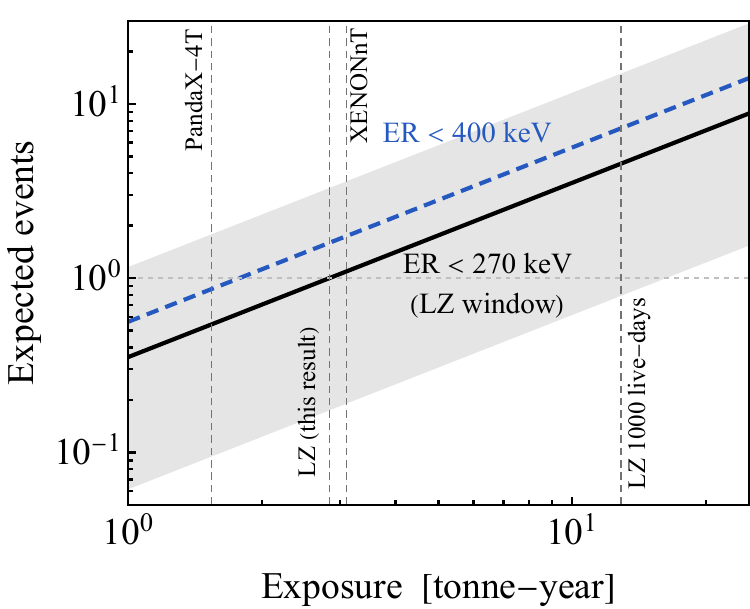}
  \caption{Expected events from $m_\chi = 42$ GeV and $|\delta|=1\MeV$ exothermic DM vs. xenon exposure, normalized to one event in LZ's $2.84$ tonne-years (solid). The shaded band shows the $68\%$ Poisson interval from a single count. The dashed blue curve extends the analysis window to $400\keV$, slightly increasing the rate due to the spectra beyond the current acceptance region. The vertical lines mark published PandaX-4T and XENONnT exposures (whose analyses do not yet test $248\keV$ recoils) and the present LZ result, along with LZ's projected $1000$ live days.}
  \label{fig:projection}
\end{figure}

\emph{Projections ---}
If the LZ event is due to DM, any model yielding one event in $2.84$ tonne-years has an expected rate in the range $0.06$--$1.2$ events per tonne-year~\cite{SM}. This has implications with regard to other direct detection experiments (see Fig.~\ref{fig:projection}). The published XENONnT ($3.1$ tonne-year~\cite{XENON:2025vwd}) and PandaX-4T ($1.54$ tonne-year~\cite{PandaX:2024qfu}) exposures correspond to $1.1$ and $0.5$ expected events, respectively. Thus while a statistically significant (i.e. 5$\sigma$) confirmation will likely require more data, both experiments could lend credence to a DM interpretation. Meanwhile, LZ's projected $1000$ live days ($\simeq13$ tonne-years) would yield $\simeq4.5$ events at the central rate, with the probability of seeing none at $3\%$. Conversely, a null high-energy exposure of $8.5$ tonne-years excludes the central rate at $95\%$. Several events would begin to resolve the peak width, measuring $\mu_{\chi N}$ and breaking the mass--splitting degeneracy outright. Lastly we note that the peak position scales as $\mu_{\chi N}/m_N$ across different nuclear targets, therefore a corresponding signal in argon or germanium would appear at a different energy, providing a unique signature of the exothermic scenario.

\emph{Summary ---}
If the lone LZ230616 event is interpreted as arising due to DM scattering, then it can be explained by exothermic inelastic dark matter, which can produce a sharp peak from a down-scattering state with mass splitting within $0.65\MeV$ of $m_N\ER/\mu_{\chi N}$. The empty high-energy sideband prefers a sharper peak over broad alternatives, bounding $\mchi \lesssim 90\GeV$. Constraints on the excited-state lifetime can be satisfied for a $\gtrsim40\GeV$ state with a dark photon mediator or a lighter state with a leptophobic mediator. Such a scenario is testable on the timescale of current experiments, including with existing XENONnT and PandaX-4T data analyzed at high energy, the full 1000 day LZ exposure, or upcoming results from argon. 

\begin{acknowledgments}

\emph{Acknowledgments ---} We are grateful for helpful discussions with Nicole Bell, Jason Kumar and Nicholas Rodd. JBD acknowledges support
from the National Science Foundation under grant no. PHY2412995. JLN is supported by the Australian Research Council through the ARC Centre of Excellence for Dark Matter Particle Physics, CE200100008. The authors used Claude Fable 5 for code assistance and Fable 5 along with ChatGPT-5.6 Sol as proofreading aids. The authors take full responsibility for the content of this work, up to and including any semi-colons and em dashes.
\end{acknowledgments}

\bibliography{exothermicLZ}

\clearpage
\onecolumngrid
\section{Supplemental Material for ``Exothermic and Endothermic  Inelastic Dark Matter Interpretations at LZ: Sideband Constraints and Future Prospects''}

\subsection{Rate and halo model}

The differential nuclear recoil rate per unit detector mass is
\begin{equation}
  \frac{dR}{d\ER} =
  \frac{f_*\rho_{\rm DM}\, \sigma_p}{2 \mchi \mu_{\chi p}^2 \langle m_N\rangle}
  \sum_i x_i\, m_{N,i} A_i^2 F_i^2(\ER)\, \eta\!\left(\vmin^{(i)}\right),
  \label{eq:rate}
\end{equation}
where $\sigma_p$ is the spin-independent DM--nucleon cross section (isoscalar couplings are assumed), $f_*$ the fraction of the local density $\rho_{\rm DM} = 0.3\GeV/\mathrm{cm}^3$~residing in the excited state, and $\eta(\vmin) = \int_{v>\vmin} f(v)/v\, d^3v$ the mean inverse speed for a truncated Maxwellian with $v_0 = 238\,\mathrm{km/s}$, $v_{\rm esc} = 544\,\mathrm{km/s}$ and mean laboratory speed $v_{\rm lab} = 250\,\mathrm{km/s}$, following the conventions of Ref.~\cite{Baxter:2021pqo} used by LZ. The sum runs over the nine stable xenon isotopes with number fractions $x_i$, and $\langle m_N \rangle = \sum_i x_i m_{N,i}$ converts the per-nucleus rate to a per-unit-mass rate.

At $\ER = 250\keV$ the momentum transfer is $q \simeq 1.2\,\mathrm{fm}^{-1}$, and so the recoil lies well beyond the first zero of the Helm form factor~\cite{Helm:1956zz}, near $\ER \simeq 105\keV$, where $F^2 \sim 10^{-4}$ and varies rapidly with both $\ER$ and $A$. The form factor therefore provides a significant $\mathcal{O}(10^{-4})$ suppression that can drastically change the fitted spectra (making isotope averaging especially important). It also has large uncertainties in this region and the results should be interpreted with this in mind.

\subsection{Detector model and likelihood}

We take an exposure of $2.84$ tonne-years and model the published nuclear recoil efficiency as a $96\%$ plateau between $14$ and $250\keV$, with error functions passing through $50\%$ at $5.4$ and $269.9\keV$ at the lower and upper edges, respectively~\cite{LZ:2026axp}. The energy resolution is assumed to be a Gaussian with $\sigma_E = 23\keV$ at the event energy (taken from LZ230616's statistical uncertainty). 

Given the background LZ expectation of $0.0106\pm0.0008$ events, we set the background to zero. The unbinned likelihood for one observed event at $\ER^{1}$ is then given by
\begin{equation}
  \mathcal{L} = e^{-N_{\rm exp}}\, MT\, \epsilon(\ER^{1})\,\frac{dR}{d\ER},
\end{equation}
where $MT$ is the exposure, $\epsilon(E_R)$ is the efficiency at $E_R$, and the expected number of events is $N_{\rm exp} = MT\sigma_p\int \epsilon \frac{dR}{d\ER}\, d\ER$. Since $\sigma_p$ enters only as a normalization we profile it, setting the maximum likelihood to: $\hat\sigma_p = 1/N_{\rm exp}$ (evaluated at $\sigma_p=1$), leaving a purely shape likelihood with parameters $(\mchi,\delta)$. To derive confidence intervals we use the asymptotic $\chi^2$ distribution~\cite{Cowan:2010js}, and not that this approximation is imperfect at low statistics.

\subsection{The high-energy sideband}
\label{sec:sideband}

To validate their multiple-scintillation single-ionization background model, LZ defines a high-energy sideband at $800 < \mathrm{S1c} < 1700$~phd. Reading the recoil-energy contours of their Figs.~S4 and S6, this corresponds to roughly $350$--$680\keV$, and their Table~S6 reports zero observed events there in the science sample of the $4.71$~tonne fiducial volume.

This region is used for background validation, and is not part of the signal region. LZ do not quote a signal efficiency for this region. Our analysis is therefore based on a few assumptions. First, that the $96\%$ analysis-cut efficiency of their Fig.~S2 remains constant to higher energies. The second assumption is that the $E_R$ scale can be extrapolated above $\sim330\keV$, beyond the AmBe calibration range. Finally, we have assumed that the $\mathrm{S2c} < 10^{4.3}$ ceiling of the sideband still contains the nuclear recoil band, which appears to be approximately true from Fig.~2 of Ref.~\cite{LZ:2026axp}.

With these caveats, the sideband provides a strong constraint on the models with their peak sitting above the LZ search window. At the search-window best fits, the expected sideband yield per observed event is $0.02$ at $\mchi = 10\GeV$, $19$ at $42\GeV$, and $73$ at $100\GeV$. Treating the sideband as a second acceptance region in the same background-free likelihood gives the bounds quoted in the main text. This conclusion is robust to small errors in the assumed sideband efficiency, because the over-prediction is one to two orders of magnitude. The interval $600 < \mathrm{S1c} < 800$~phd ($\simeq 270$--$350\keV$) lies between the two published selections and is not included in either analysis. The heavier solutions tend to concentrate their peaks in this gap, and ideally this region would also be included in future analyses. The earlier LZ NREFT search ($0.9$ tonne-year over the same $\mathrm{S1c}$ range) saw no high-energy excess~\cite{LZ:2023lvz}. Treated as an additional exposure it would further reduce the preferred rate. We emphasize that the present analysis is no substitute for a collaboration analysis, with a properly calibrated efficiency, energy scale and background model in this region.

\subsection{Exposure statistics}

For one observed count the central $68\%$ interval on the Poisson mean is $[0.174,\,3.29]$; dividing by $2.84$ tonne-years gives a rate of $0.35^{+0.81}_{-0.29}$ events per tonne-year, the band shown in Fig.~3 of the main text. A background-free exposure with zero events of $\mathcal{E}_{95} \simeq 8.5$ tonne-years rejects the central rate at $95\%$. LZ's projected $1000$ live days at the present fiducial mass corresponds to $12.9$ tonne-years and $4.5$ expected events at the central rate. The Poisson probability of observing no additional events beyond the current exposure is $e^{-3.5} \simeq 3\%$.

\end{document}